\documentclass[twocolumn,showpacs,preprintnumbers,amsmath,amssymb,superscriptaddress]{revtex4-1}
\usepackage{graphicx}% Include figure files
\usepackage{dcolumn}% Align table columns on decimal point
\usepackage{bm}% bold math
\usepackage{tabularx}
\usepackage{makecell}
\usepackage{braket}
\usepackage{lipsum}
\usepackage{ulem}
\usepackage{amsmath,amssymb} 
\newcolumntype{M}{>{\centering\arraybackslash}m{1.85cm}}
\usepackage[export]{adjustbox}
\usepackage{longtable}
\usepackage{float}
\usepackage{xcolor}
\usepackage{color}   %May be necessary if you want to color links
\usepackage{hyperref}
\hypersetup{
	colorlinks=true, %set true if you want colored links
	linktoc=all,     %set to all if you want both sections and subsections linked
	linkcolor=blue,  %choose some color if you want links to stand out
}

\makeatletter
\newcommand{\colorcaption}[2][]{%
	\begingroup%
	\renewcommand{\@caption@fignum@sep}{ (Color online). }%
	\caption[#1]{#2}%
	\endgroup%
}
\makeatother
\newcommand{\orcid}[1]{\href{https://orcid.org/#1}{\hskip2pt\includegraphics[width=9pt]{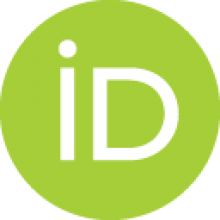}}}
\begin{document}
% !TeX spellcheck = en_US

%\begin{document}
	
 \title{\color{black}Modification of single-hole-like states by configuration mixing in the $^{99-131}$In}
	
	\author{ Deepak Patel\orcid{0000-0002-7669-1907}}
	\address{Department of Physics, Indian Institute of Technology Roorkee, Roorkee 247667, India}
	\author{Praveen C. Srivastava\orcid{0000-0001-8719-1548}}
	\email{Contact author: praveen.srivastava@ph.iitr.ac.in}
	\address{Department of Physics, Indian Institute of Technology Roorkee, Roorkee 247667, India}
	\author{Noritaka Shimizu\orcid{0000-0003-1638-1066}}
	\address{Center for Computational Sciences, University of Tsukuba, 1-1-1, Tennodai Tsukuba,\\ Ibaraki 305-8577, Japan}
	\author{Yutaka Utsuno\orcid{0000-0001-5404-7320}}
    \email{Contact author: utsuno.yutaka@jaea.go.jp}
	\address{Advanced Science Research Center, Japan Atomic Energy Agency, Tokai, Ibaraki 319-1195, Japan}
	\address{Center for Nuclear Study, University of Tokyo, 7-3-1 Hongo, Bunkyo-ku, \\Tokyo 113-0033, Japan}

	\date{\hfill \today}
	
	%%%%%%%%%%%%%%%%%%%%%%%%%%%%%%%%%%%%%%%%%
	
	\begin{abstract}

  Large-scale shell-model calculations are performed for the $9/2^+_{\rm g.s.}$, $1/2^-_1$, $3/2^-_1$, and $5/2^-_1$ states in the odd-$A$ indium isotopes with $N=50-82$. The calculated energy levels, electromagnetic moments, and spectroscopic factors exhibit remarkable agreement with the experimental data due to significant configuration mixing for the neutron numbers away from the closed shells. {\color{black}The $1/2^-_1$ energy levels closely follow the trend of effective single-particle energies, which are determined using the fractional occupancies of neutron orbitals. However, configuration mixing with the proton $p_{3/2}$ and $f_{5/2}$ orbitals in the actual shell-model calculations plays a crucial role in accurately reproducing the positions of the $1/2^-_1$ levels, ensuring better agreement with the experimental data across the entire isotopic chain.}

	\end{abstract}

	
	\maketitle
	%%%%%%%%%%%%%%%%%%%%%%%%%%%%%%%%%%%%%%%%%%%%%%%%%%%%%%%%%%%
	
	\section{Introduction}

	{\color{black}The structure of nuclei lying in the tin region provides a unique testing ground for investigating shell evolution due to the presence of the proton magic number $Z=50$ \cite{Jones, Gorges, Morris, Stone, Gorska, Nesterenko, Hakala, Vernon_nature, Yordanov, Xu, Chen, Spieker, Yordanov1, Kankainen}, which allows for the study of the different isotopic chains between two neutron shell closures, $N=50$ and 82. The study of isotopes in this region has significantly contributed to our understanding of the evolution of shell structure \cite{Otsuka1}, which is influenced by nuclear forces.} In particular, the isotopes that lie in the neighborhood of tin, such as indium (In) and antimony (Sb), provide more direct insights into the shell structure due to the occurrence of a single proton hole (particle) outside the proton magic number $Z=50$.
	
	{\color{black}For Sb isotopes, Schiffer $et$ $al.$ \cite{Schiffer} observed very strong peaks of the $11/2^-_1$ and $7/2^+_1$ states in the $^{A-1}$Sn($\alpha$, $t$)$^A$Sb reactions. Regarding these two levels as proton $h_{11/2}$ and $g_{7/2}$ single-particle states, respectively, the authors of Ref. \cite{Schiffer} proposed that the spin-orbit interaction strongly changes with neutron excess on the basis of the sharp change of the level splitting of the $11/2^-_1$ and $7/2^+_1$ states. This phenomenon was later interpreted by Otsuka $et$ $al.$ \cite{Otsuka2} as evidence of tensor-force-driven shell evolution. However, in another study, Sorlin and Porquet \cite{Sorlin} proposed an alternative explanation, attributing this evolution to the coupling of the core excitation: the $11/2^-_1$ are strongly mixed with the $\pi d_{5/2} \times 3^-_1$ and $\pi g_{7/2} \times 3^-_1$ core excited states. This difference in interpretation is further complicated by different many-body nuclear-structure calculations, such as those by Utsuno $et$ $al.$ \cite{Utsuno} or Afanasjev $et$ $al.$ \cite{Afanasjev}, which support the conclusion that the observed level evolution is driven by the tensor-force or the particle-core coupling mechanism, respectively. It is most likely that the $11/2^-_1$ and $7/2^+_1$ states are not either pure single-particle or pure core excited states but are strongly mixed ones. Hence, precisely evaluating configuration mixing in these states is also highly desired for deducing shell evolution.
		
		A similar situation is expected to occur in In isotopes with one proton less than Sn isotopes. Large-scale shell-model calculations provide a robust approach \cite{Caurier} to address this issue.} Such studies offer insight into the role of monopole interactions and configuration mixing in shaping nuclear structure properties across isotopic chains. While cadmium isotopes have been extensively studied to understand the impact of two proton holes relative to the tin core \cite{Zuker, Patel, Yordanov2, Siciliano}, indium isotopes, with a single proton hole at $Z=49$, provide a distinct vantage point for examining the evolution of {\color{black}proton-hole-like} states and their interaction with various neutron configurations. Investigating the structural evolution across indium isotopes may yield a more unified understanding of shell evolution in this region, which could reconcile the observed data from both Sb and In isotopes in terms of the driving forces behind the evolution of nuclear structure.

      The present work aims to systematically describe the proton-hole-like levels in the In isotopes {\color{black}using large-scale shell-model calculations in order to deduce the configuration mixing in these states across the entire isotopic chain, which drives their evolution.} We utilize the same proton-neutron interaction that was employed for the study of Sb isotopes \cite{Utsuno} on the basis of the $V_{\rm MU}$ interaction \cite{Otsuka} to provide a consistent description of In isotopes. For this purpose, nuclear energy levels, spectroscopic factors, and nuclear moments in indium isotopes are calculated and compared to the experimental data. In particular, nuclear moments and spectroscopic factors \cite{Vernon_nature, Weiffenbach} serve as sensitive measures of configuration mixing, providing valuable insights into the {\color{black}underlying mechanisms driving} the structural evolution of these isotopes. We examine the impact of this evolution on the single-proton-hole states $9/2^+_{\rm g.s.}$, $1/2^-_1$, $3/2^-_1$, and $5/2^-_1$, especially for the $1/2^-_1$ state, which was not well understood in previous studies. {\color{black}Furthermore, investigating the influence of central and tensor forces by the monopole matrix element on the energy variations of these single-hole-like states as a function of neutron numbers is a key aspect of this study.}

	This paper is organized as follows: In Sec. \ref{section2}, we briefly show the outline of our calculations. In Sec. \ref{section3}, we compare the calculated single-hole-like levels in $^{99-131}$In to the experimental data and probe the role of configuration mixing through electromagnetic moments and spectroscopic factors. Furthermore, the systematics of the $1/2^-_1$ excitation energies {\color{black}are discussed in terms of effective proton single-hole energies} and correlation energies. Finally, we summarize our results and conclude the paper in Sec. \ref{section4}.
 %\vspace{-4mm}

	\section{Formalism} \label{section2}
	
	The shell-model Hamiltonian $H$ consists of single-particle energies $T$ and two-body {\color{black}interaction} terms $V$, 
\begin{equation}
    H = T+V, 
\end{equation}
with 
\begin{gather}
    T = \sum_{\alpha}\epsilon^0_{\alpha}\hat{N}_\alpha, \\
        V = \sum_{\alpha \le \beta,  \gamma \le \delta JM}
  %\langle j_\alpha j_\beta |V|j_\gamma j_{\delta} \rangle_{J} 
  V_J(j_{\alpha} j_{\beta} j_{\gamma} j_{\delta})
		C_{j_{\alpha}j_{\beta}JM}^{\dagger}C_{j_{\gamma}j_{\delta}JM}. 
\label{twobody}
\end{gather}
The labels $\alpha$ to $\delta$ stand for the set of quantum numbers, $\alpha=\{\rho_{\alpha}n_{\alpha}l_{\alpha}j_{\alpha}\}$, where $\rho=\pi$ for proton or $\rho=\nu$ for neutron. Instead of $\alpha$ etc., $j_{\alpha}$ is often used to label the orbitals as expressed in Eq. (\ref{twobody}).  $\hat{N}_{\alpha}=\sum_{m_{\alpha}}c^{\dagger}_{{\alpha}m_{\alpha}}c_{{\alpha}m_{\alpha}}$ is the particle number operator, and $\epsilon^0_{\alpha}$ is the single-particle energy on top of the inert core employed. $C^{\dagger}_{j_{\alpha}j_{\beta}JM}$ is the pair creation operator, 
\begin{equation}
    C^{\dagger}_{j_{\alpha}j_{\beta}JM} = \mathcal{N}_{\alpha \beta} \sum_{m_{\alpha} m_{\beta}} 
    (j_{\alpha} m_{\alpha} j_{\beta} m_{\beta} | JM) c^{\dagger}_{j_{\alpha} m_{\alpha} } c^{\dagger}_{j_{\beta} m_{\beta}},
\end{equation}
with the normalization factor $\mathcal{N}_{\alpha \beta}=(1+\delta_{\alpha\beta})^{-1/2}$, and $C_{j_{\alpha}j_{\beta}JM}$ is the Hermitian conjugate of $C^{\dagger}_{j_{\alpha}j_{\beta}JM}$. The symbol $V_J(j_{\alpha} j_{\beta} j_{\gamma} j_{\delta})$ denotes the antisymmetrized two-body matrix elements, $\langle j_{\alpha} j_{\beta}JM | V | j_{\gamma} j_{\delta}JM \rangle$, where $| j_{\alpha} j_{\beta}JM \rangle$ stands for $C^{\dagger}_{j_{\alpha}j_{\beta}JM} | - \rangle$ with the vacuum {\color{black}state} $|-\rangle$. Note that the value of $\langle j_{\alpha} j_{\beta}JM | V | j_{\gamma} j_{\delta}JM \rangle$ does not depend on $M$, and we omit $M$ in $V_J(j_{\alpha} j_{\beta} j_{\gamma} j_{\delta})$. 
	
    In the present study, we performed shell-model calculations for the odd-$A$ $^{99-131}$In isotopes. The valence shell consists of four proton orbitals ($0f_{5/2}$, $1p_{3/2}$, $1p_{1/2}$, and $0g_{9/2}$) and five neutron orbitals ($1d_{5/2}$, $2s_{1/2}$, $1d_{3/2}$, $0g_{7/2}$, and $0h_{11/2}$) with the inert core $^{78}$Ni. The neutron-neutron interaction was taken from the SNBG1 interaction \cite{SNBG1}, whose matrix elements were semi-empirically determined to reproduce the energy levels of Sn isotopes in the neutron valence shell consisting of ($1d_{5/2}$, $2s_{1/2}$, $1d_{3/2}$, $0g_{7/2}$, and $0h_{11/2}$) orbits. The proton-proton interaction is omitted for simplicity because it does not affect the indium isotopes with a proton hole in the valence shell. 
    The proton-neutron interaction was taken from a version of the $V_{\rm MU}$ interaction that was employed in the SDPF-MU interaction for the $sd$-$pf$ shell \cite{sdpfmu}, in which a two-body spin-orbit interaction was also included. This interaction was also used as the proton-neutron interaction for the study of Sb isotopes \cite{Utsuno} with the overall central force scaled by 0.84 to reproduce the binding energies of Sn isotopes. This overall scaling was {\color{black}also introduced} in the present study for a consistent description of In isotopes. All the two-body matrix elements were scaled by $A^{-0.3}$ following the recipe of the SNBG1 interaction \cite{SNBG1}. The single-particle energies of $0f_{5/2}$, $1p_{3/2}$, $1p_{1/2}$, and $0g_{9/2}$ proton orbitals were adjusted so that the corresponding single-hole states in $^{131}$In reported in Ref. \cite{Vaquero} could be reproduced. {\color{black}However, the single-particle energies of the neutron orbitals were adjusted to reproduce the energy levels in $^{101}$Sn as predicted by the SNBG1 interaction \cite{SNBG1}.} We {\color{black}also} validated the strength of the {\color{black}employed interaction in this study} by successfully reproducing the experimental proton separation energy difference between $^{132}$Sn and $^{104}$Sn.

    For our calculations, the shell-model code kshell \cite{KShell} has been used to diagonalize the shell-model Hamiltonian matrices. We have performed our calculations without any truncation in the present model space. The largest $M$-scheme dimension is $3.4 \times 10^8$ for $^{115}$In.

        \begin{figure*}
		\includegraphics[width=155mm]{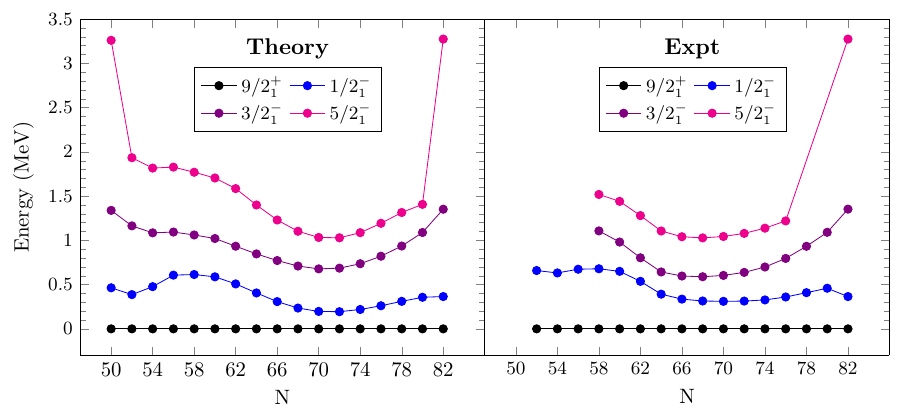}
		\caption{Comparison of the shell-model energy levels for $9/2^+_{1}$, $1/2^-_1$, $3/2^-_1$, and $5/2^-_1$ states in odd-$A$ indium isotopes with the experimental data \cite{Xu, NNDC, Vaquero}.}
		\label{low_spin_states}
	\end{figure*}

	\section{Results and Discussions} \label{section3}

	  In Sec. \ref{sec:levels}, we first show the systematics of possible proton-hole-like states ($9/2^+_{\rm g.s.}$, $1/2^-_1$, $3/2^-_1$, and $5/2^-_1$) in $^{99-131}$In and probe their configuration mixing from seniority distribution. In Sec. \ref{sec:moments}, the obtained configuration mixing is validated with electromagnetic moments. In Sec. \ref{sec:sfactors}, we decompose the calculated $9/2^+_{\rm g.s.}$ and $1/2^-_1$ wave functions into the major configurations that are classified according to the hierarchy of energy. The depletion of the single-hole strength is quantified by the spectroscopic factors. {\color{black}In Sec. \ref{sec:espe}, the systematics of the $1/2^-_1$ excitation energies is compared with the effective single-hole energies of the proton orbital $p_{1/2}$, and their behavior is explained based on the influence of the central and tensor forces through the monopole matrix element.}

\subsection{Systematics of the $9/2^+_{\rm g.s.}$, $1/2^-_1$, $3/2^-_1$, and $5/2^-_1$ levels in indium isotopes}
\label{sec:levels}

        \begin{figure}
		\includegraphics[width=88mm]{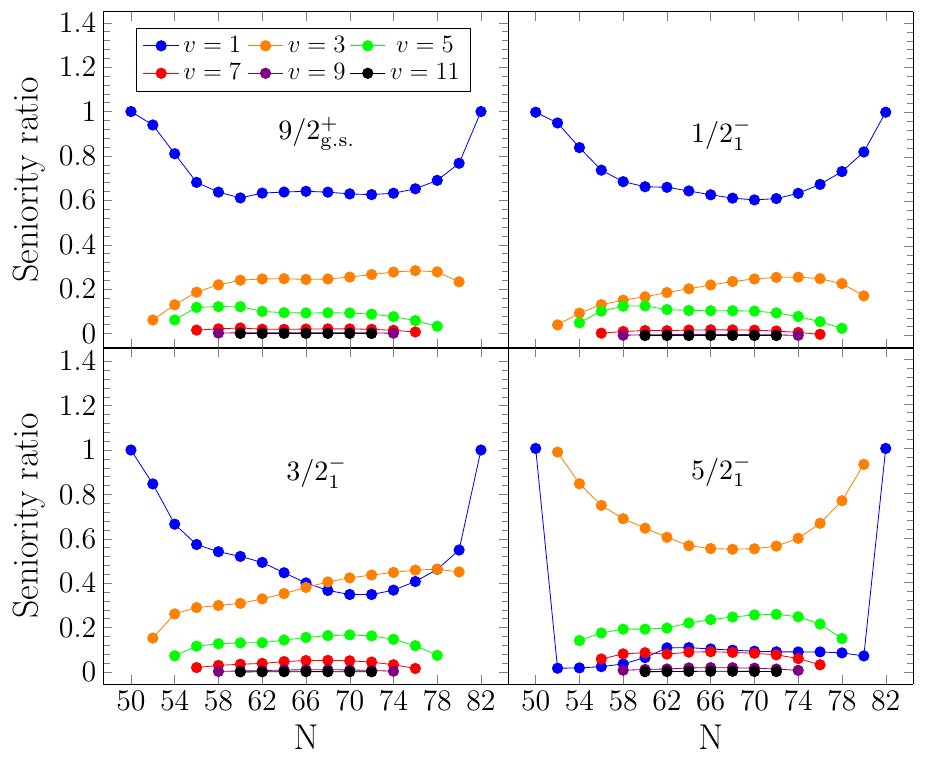}
		\caption{Seniority of the $9/2^+_{\rm g.s.}$, $1/2^-_1$, $3/2^-_1$, and $5/2^-_1$ states in odd-$A$ indium isotopes.}
		\label{Seniority}
	\end{figure}

   \begin{figure}
        	\includegraphics[width=85mm]{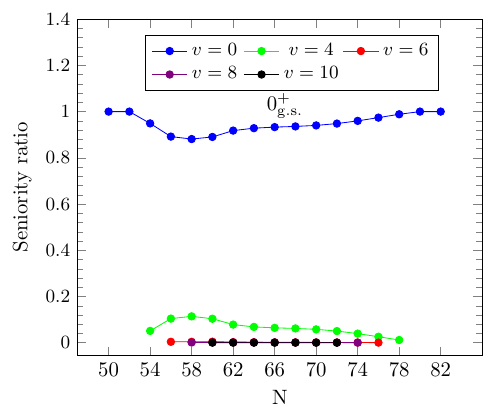}
        	\caption{Seniority of the $0^+_{\rm g.s.}$ state in even-$A$ tin isotopes.}
        	\label{Seniority_Sn}
        \end{figure}

  In Fig. \ref{low_spin_states}, we compare the calculated energy levels of the $9/2^+_1$, $1/2^-_1$, $3/2^-_1$, and $5/2^-_1$ states with the available experimental data \cite{Xu, NNDC, Vaquero} across the odd-$A$ indium isotopic chain. To the best of our knowledge, for the first time in the present work, the experimental trend of $1/2^-_1$ states throughout the indium chain has been successfully reproduced by shell-model calculations. The comparison in Fig. \ref{low_spin_states} {\color{black}shows} a generally consistent match between the calculated and experimental trends. For $^{107,109,113,123,125}$In, firm spin-parity assignments of the $5/2^-$ levels are not provided, and we propose assignments based on the shell-model predictions, as shown in Fig. \ref{low_spin_states}. For $^{107}$In, the shell-model calculation predicts the $5/2^-_1$ state at 1770 keV. Experimentally, there is an unconfirmed spin-parity state $(1/2,3/2,5/2)^-$ at 1519 keV, and no calculated $1/2^-$ or $3/2^-$ states lie in this energy region, leading us to assign the $5/2^-_1$ state at 1519 keV. Similarly, we have considered that the experimental $5/2^-$ state at 1441 keV in $^{109}$In, the $3/2^-,5/2^-$ state at 1106 keV in $^{113}$In, the $(3/2^-,5/2^-)$ state at 1138 keV in $^{123}$In, and the $(1/2,3/2,5/2^-)$ state at 1220 keV in $^{125}$In could be the $5/2^-_1$ states.
  
  The seniority number ($v$) provides useful information on nuclear structure, particularly for examining the single-particle structure of a state. It is defined as the unpaired particles, which are not pairwise coupled to the angular momentum $J=0$ \cite{Racah, Flowers}. In Fig. \ref{Seniority}, we illustrate the seniority distribution for the $9/2^+_{\rm g.s.}$, $1/2^-_1$, $3/2^-_1$, and $5/2^-_1$ states across the odd-$A$ indium isotopic chain, highlighting the predominant seniority values for each state. The $9/2^+_{\rm g.s.}$ and $1/2^-_1$ states are dominated by $v=1$ throughout the isotope chain, thus indicating that the majority of the wave function is composed of $\pi(g_{9/2})$ and $\pi(p_{1/2})$ holes, respectively, on top of the ground states of the corresponding tin isotopes. For the $3/2^-_1$ and $5/2^-_1$ states, we observe a different structural behavior. The $3/2^-_1$ state retains single-particle character across most indium isotopes but shows deviations for isotopes in the after mid-shell region, particularly in the range $N=68-76$, where fragmentation in the wave function leads to the evolution in the collectivity. These deviations are also reflected in the spectroscopic factors discussed later, which suggest a reduced single-particle dominance for this state. On the other hand, the $5/2^-_1$ state displays a pronounced departure from the single-particle characteristics, except for the neutron shell closures, at $N=50$ and $N=82$. This occurs because the energy of a $\pi(p_{1/2})$ hole coupled to the $2^+_1$ state in the corresponding tin isotope is lower than a $\pi(f_{5/2})$ hole coupled to the $0^+_{\rm g.s.}$ state for $52 \le N \le 80$.

  In Fig. \ref{Seniority_Sn}, the seniority distributions of the ground states of tin isotopes are presented to compare those of the $9/2^+_{\rm g.s.}$, $1/2^-_1$, $3/2^-_1$, and $5/2^-_1$ states in indium isotopes shown in Fig. \ref{Seniority}. For the tin isotopes, the lowest seniority number ($v=0$) accounts for $\gtrsim 90$\% of the ground state, whereas the corresponding number in the indium isotopes ($v=1$) is significantly reduced down to $\approx 60$\% even for the $9/2^+_{\rm g.s.}$ and the $1/2^-_1$ levels. Since the seniority numbers for protons must always be $v=1$ in the present model space, this reduction indicates additional configuration mixing driven by a proton hole.

	\begin{table*}
		\centering
		\caption{Electric quadrupole and magnetic moments of odd-$A$ indium isotopes. In our {\color{black}calculations,} we have taken two sets of effective charges $(e_p, e_n)=(1.6, 0.8)e$ \cite{Honma} and {\color{black}$(1.65, 1.04)e$}; the gyromagnetic ratios for the spin angular momenta are taken as $g_s^p=3.910$ and $g_s^n=-2.678$, and those for the orbital angular momenta are $g_l^p=1.0$ and $g_l^n=0.0$ \cite{Patel}.}
		\begin{ruledtabular}
			\begin{tabular}{lccccccccc}
				
				%\hline \hline
				&          & \multicolumn{4}{c}{$Q$ ($e$b)}    & \multicolumn{3}{c}{$\mu$ ($\mu_N$)} \\ % \T\B\\
				\cline{3-6}
				\cline{7-9}
				
				~$A$  & $J^{\pi}$ & \multicolumn{2}{c}{Theory} &  Expt.  & Expt. \cite{Vernon} & Theory & Expt.  &  Expt. \cite{Vernon}	\\ % \T\B\\

                \cline{3-4}

                    &  & $(1.6, 0.8)e$ & {\color{black}$(1.65, 1.04)e$} &  &  &  &  & \\
                   
				\hline
				99   & $9/2^+_{1}$  & 0.310 & {\color{black}0.320} & - & -  &  5.955  & - & - \\
				
				& $1/2^-_{1}$  & - & - & - & -  &  0.015  & - & - \\
				
				101  & $9/2^+_{1}$  & 0.377 & {\color{black}0.408} & {\color{black}0.486(16) \cite{Karthein}} & -  &  5.875  & {\color{black}5.861(10) \cite{Karthein}} & - \\

                         & $1/2^-_{1}$  & - & - & - & -  &  $-0.022$  & $-0.113(5)$ \cite{Karthein} & - \\
				
				103  & $9/2^+_{1}$  & 0.474 & {\color{black}0.537} & {\color{black}0.671(17) \cite{Karthein}}  & -  &  5.804  & {\color{black}5.760(3) \cite{Karthein}} & - \\

                         & $1/2^-_{1}$  & - & - & - & -  &  $-0.079$ & {\color{black}$-0.125(4)$ \cite{Karthein}} & - \\
				
				105  & $9/2^+_{1}$  & 0.573 & {\color{black}0.667} & 0.79(5) \cite{NDS} & -  & 5.735 & 5.667(5) \cite{NDS} & - \\

                         & $1/2^-_{1}$  & - & - & - & -  &  $-0.120$  & {\color{black}$-0.144(1)$ \cite{Karthein}} & - \\
				
				107  & $9/2^+_{1}$  & 0.627 & {\color{black}0.740} & 0.77(5) \cite{NDS} & -  & 5.678 & 5.577(8) \cite{NDS} & - \\

                         & $1/2^-_{1}$  & - & - & - & -  &  $-0.139$  & {\color{black}$-0.154(3)$ \cite{Karthein}} & - \\
				
				109  & $9/2^+_{1}$  & 0.661 & {\color{black}0.784} & 0.80(3) \cite{NDS} & -  & 5.627 & 5.530(4) \cite{NDS} & - \\

                         & $1/2^-_{1}$  & - & - & - & -  &  $-0.158$  & {\color{black}$-0.171(1)$ \cite{Karthein}} & - \\
				
				111  & $9/2^+_{1}$  & 0.661 & {\color{black}0.784} & 0.76(2) \cite{NDS} & -  & 5.611 & 5.495(7) \cite{NDS} & - \\

                         & $1/2^-_{1}$  & - & - & - & -  &  $-0.180$  & {\color{black}$-0.185(1)$ \cite{Karthein}} & - \\
				
				113  & $9/2^+_{1}$  & 0.665 & {\color{black}0.789} & 0.761(5) \cite{NDS} & 0.767(27) & 5.587 & 5.5208(4) \cite{NDS} & 5.5264(19) \\
				
				& $1/2^-_{1}$  & - & - & - & - & $-0.201$ & $-0.21043(3)$ \cite{NDS} & $-0.21(1)$ \\
				
				115  & $9/2^+_{1}$  & 0.669 & {\color{black}0.793} & 0.772(5) \cite{NDS} & 0.784(42) & 5.564 & 5.5326(4) \cite{NDS} & 5.541(2) \\
				
				& $1/2^-_{1}$  & - & - & - & - & $-0.220$ & $-0.24362(5)$ \cite{NDS} & $-0.2405(38)$ \\
				
				117  & $9/2^+_{1}$  & 0.674 & {\color{black}0.801} & 0.790(10) \cite{NDS} & 0.807(22) & 5.535 & 5.511(4) \cite{NDS} & 5.5286(43) \\
				
				& $1/2^-_{1}$  & - & - & - & - & $-0.237$ & $-0.25136(4)$ \cite{NDS} & $-0.2766(27)$ \\
				
				119  & $9/2^+_{1}$  & 0.679 & {\color{black}0.807} & 0.814(7) \cite{NDS} & 0.794(23) & 5.498 & 5.507(10) \cite{NDS} & 5.499(62) \\
				
				& $1/2^-_{1}$  & - & - & - & - & $-0.250$ & $-0.319(5)$ \cite{NDS} & $-0.342(12)$ \\
				
				121  & $9/2^+_{1}$  & 0.674 & {\color{black}0.800} & 0.776(10) \cite{NDS} & 0.803(23) & 5.461 & 5.494(5) \cite{NDS} & 5.575(62) \\
				
				& $1/2^-_{1}$  & - & - & - & - & $-0.260$ & $-0.354(4)$ \cite{NDS} & $-0.3600(41)$ \\
				
				123  & $9/2^+_{1}$  & 0.654 & {\color{black}0.775} & 0.722(9) \cite{NDS} & 0.736(23) & 5.429 & 5.483(7) \cite{NDS} & 5.442(61) \\
				
				& $1/2^-_{1}$  & - & - & - & - & $-0.266$ & $-0.399(4)$ \cite{NDS} & $-0.4047(54)$ \\
				
				125  & $9/2^+_{1}$  & 0.616 & {\color{black}0.724} & 0.68(3) \cite{NDS} & 0.673(24) & 5.410 & 5.494(9) \cite{NDS} & 5.496(24) \\
				
				& $1/2^-_{1}$  & - & - & - & - & $-0.263$ & $-0.432(4)$ \cite{NDS} & $-0.450(17)$ \\
				
				127  & $9/2^+_{1}$  & 0.557 & {\color{black}0.647} & 0.56(3) \cite{NDS} & 0.588(29) & 5.421 & 5.514(8) \cite{NDS} & 5.5321(14) \\
				
				& $1/2^-_{1}$  & - & - & - & - & $-0.241$ & - & $-0.4355(24)$ \\
				
				129  & $9/2^+_{1}$  & 0.470 & {\color{black}0.532} & - & 0.487(13) & 5.504 & - & 5.5961(23) \\
				
				& $1/2^-_{1}$  & - & - & - & - & $-0.167$ & - & $-0.38709(58)$ \\
				
				131  & $9/2^+_{1}$  & 0.336 & {\color{black}0.347} & - & 0.31(1)  & 5.955 & - & 6.312(14) \\
				
				& $1/2^-_{1}$  & - & - & - & - & 0.015 & - & $-0.0514(48)$ \\

			\end{tabular}
		\end{ruledtabular}
		\label{Moments}
	\end{table*}
	
\begin{table*}
	\centering
	\caption{The calculated $B(E2)$ and $B(M1)$ transitions between the  proton-hole-like states in indium isotopes compared to the experimental data \cite{NNDC_NUDAT}. The effective $E2$ and $M1$ operators are the same as those used for Table \ref{Moments}.}
	\begin{ruledtabular}
		\begin{tabular}{ccccccc}
			%\hline \hline
			
			Isotope & \makecell{Transition \\ ($J^{\pi}_i\rightarrow J^{\pi}_f$)}	& \multicolumn{3}{c}{\makecell{$B(E2)$ \\ (W.u.)}} & \multicolumn{2}{c}{\makecell{$B(M1)$ \\ (W.u.)}}	 \\
			
			\cline{3-5}
			\cline{6-7}
			
			&  & \multicolumn{2}{c}{Theory} & Expt. & Theory & Expt. \\
			
			\cline{3-4}
			
			&  &  $(1.6,0.8)e$  & {\color{black}$(1.65, 1.04)e$}  &  &  &  \\
			
			\hline
			
			&  &  &  &  &  &  \\
			
			$^{109}$In & $3/2^-_1\rightarrow 1/2^-_1$  & 12.2 & {\color{black}17.3} & - & 0.39 & $0.96^{+30}_{-31}$ \\
			
			$^{111}$In       & $5/2^-_1\rightarrow 1/2^-_1$  & 14.6 & {\color{black}22.2} & 0.47 $>$ &  &  \\
			
			$^{115}$In & $3/2^-_1\rightarrow 1/2^-_1$  & 14.5 & {\color{black}21.3} & - & 0.38 & $\ge$0.0047 \\

		\end{tabular}
	\end{ruledtabular}
	\label{BE2_BM1}
\end{table*}

		\subsection{Electromagnetic observables}
        \label{sec:moments}
	
\begin{figure}
    \includegraphics[width=87mm]{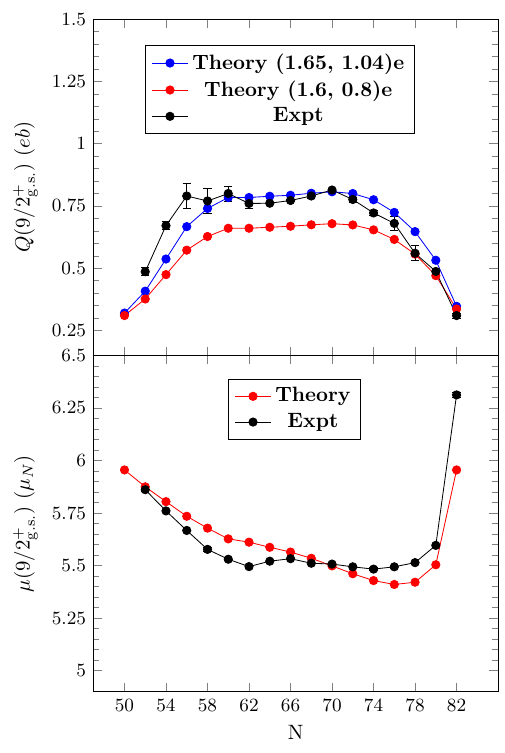}
    \caption{Comparison between the calculated and experimental \cite{NDS, Vernon} electric quadrupole (upper) and magnetic dipole (lower) moments of $9/2^+_{\rm g.s.}$ for odd-$A$ indium isotopes.}
    \label{quadrupole_magnetic}
    \end{figure}

	  Here, we discuss the electric quadrupole $(Q)$ and magnetic dipole $(\mu)$ moments of the proton-hole-like states $9/2^+_{\rm g.s.}$ and $1/2^-_1$ in the odd-$A$ indium isotopic chain. These two observables are important to probe the effect of configuration mixing and extract information about the nature of wave functions \cite{Bohr}. Here, we mainly focus on studying the systematics of electric quadrupole and magnetic moment of $9/2^+_{\rm g.s.}$; {\color{black}the comparisons of the theoretical and experimental $Q(9/2^+_{\rm g.s.})$ and $\mu(9/2^+_{\rm g.s.})$ values are depicted in Fig. \ref{quadrupole_magnetic}.} Also, the theoretical and experimental data for $Q(9/2^+_{\rm g.s.})$, $\mu(9/2^+_{\rm g.s.})$, and $\mu(1/2^-_1)$ values are reported in Table \ref{Moments}.

  We consider two distinct sets of  effective charges $(e_p, e_n)$, as $(1.6, 0.8)e$ and {\color{black}$(1.65, 1.04)e$}. The first set has been used in previous studies \cite{Honma, Patel}, while the latter is determined through a chi-square fitting method. In this procedure, the neutron effective charge ($1.04e$) is first derived using the theoretical and corresponding experimental $B(E2; 2^+_1 \rightarrow 0^+_{\rm g.s.})$ transition strengths in Sn isotopes. Subsequently, with the neutron effective charge fixed, the optimal proton effective charge {\color{black}($1.65e$)} has been obtained utilizing the theoretical and experimental quadrupole moments of the $9/2^+_{\rm g.s.}$ in {\color{black}the} indium isotopes. The shell-model calculations predict positive $Q$ values for $9/2^+_{\rm g.s.}$ across the $^{99-131}$In chain, demonstrating a consistent trend as the neutron number increases. The gradual change in $Q$ values reflects structural evolution with neutron number due to significant configuration mixing in the mid-shell region, while the decrement near neutron shell closures indicates changes from the collective behavior to the single-particle nature and clear reduction in the nuclear charge polarization \cite{Vernon_nature}. The overall agreement between theoretical and experimental $Q(9/2^+_{\rm g.s.})$ values, especially with effective charges {\color{black}$(1.65, 1.04)e$} (see upper panel of Fig. \ref{quadrupole_magnetic}), indicates that the shell-model interaction used in these calculations effectively captures the primary characteristics of the $9/2^+_{\rm g.s.}$ in the indium isotopes.

    In the lower panel of Fig. \ref{quadrupole_magnetic} we have plotted the shell-model predicted and experimental magnetic moment $\mu(9/2^+_{\rm g.s.})$ in $^{99-131}$In for comparison. The effective $g$ factors we take are $(g_l^p,g_l^n)=(1.0,0.0)$ and $(g_s^p,g_s^n)=(3.910, -2.678)$ for the orbital and spin angular momenta, respectively, following Ref. \cite{Patel}.
    The experimental magnetic moments are quite stable except at $N=82$. The present calculations well reproduce this property, but show a shallow minimum at $N=76$ in contrast to the experimental data. We reasonably reproduce the abrupt increment in $\mu(9/2^+_{\rm g.s.})$ at $N=82$, whereas the difference between the $N=82$ and $N=80$ results is not completely reproduced. As indicated by the seniority variation (see Fig. \ref{Seniority}), the proton hole drives additional configuration mixing for $9/2^+_{\rm g.s.}$ in the indium isotopes away from {\color{black}neutron} shell closures, leading to a reduced $\mu(9/2^+_{\rm g.s.})$ value. A detailed analysis of the configuration mixing is discussed later.

	Further, the observed magnetic moment of $1/2^-_1$ state decreases with the increase of neutron number up to $^{125}$In, and beyond that it increases in the $^{127-131}$In isotopes. Shell-model calculated $\mu$ moments for $1/2^-_1$ states are slightly higher than the observed values but show almost the same trend as in the experimental data.
	
	Similarly to the effective charges, effective $g$ factors can be determined by the chi-square fitting. The resulting agreement with the experimental data is, however, not significantly improved, and the obtained {\color{black}$g$ factors} deviate from the standard values. {\color{black}To avoid such overfitting, we have presented the calculated magnetic moments in Fig. \ref{quadrupole_magnetic} and Table \ref{Moments} only with the $g$ factors $(g_l^p,g_l^n)=(1.0,0.0)$ and $(g_s^p,g_s^n)=(3.910, -2.678)$, which are in accordance with the systematics.}
	
	We have also calculated the root mean square (rms) deviation in both observables ($Q$ and $\mu$) using the formula
	
	\begin{equation}\label{eq4}
		\text{rms}=\sqrt{\frac{1}{N}\sum_{k=1}^{N}(W^k_{expt}-W^k_{th})^2}.
	\end{equation}
	
   Here, $W^k_{expt}$ and $W^k_{th}$ denote the experimental and theoretical observables, respectively. In the calculation of rms deviation, the experimental data for $Q(9/2^+_{\rm g.s.})$ and $\mu(9/2^+_{\rm g.s.})$ in $^{129,131}$In and for $\mu(1/2^-_1)$ in $^{127-131}$In are taken from Ref. \cite{Vernon}, while the experimental $Q(9/2^+_{\rm g.s.})$ and $\mu(9/2^+_{\rm g.s.})$ values for $^{101,103}$In and the $\mu(1/2^-_1)$ values for $^{101-111}$In are taken from Ref. \cite{Karthein}. All other experimental data for $Q$ and $\mu$ values are taken from Ref. \cite{NDS}. By considering the maximum experimental uncertainties in $Q$ moments, the estimated rms deviations corresponding to the effective charges $(1.6,0.8)e$ and $(1.65,1.04)e$ are $0.117^{+0.020}_{-0.018}$ and $0.061^{+0.009}_{-0.001}$ eb, respectively. Similarly, our computed rms deviation for $\mu$ moments is $0.10669^{+0.00044}_{-0.00003}$ $\mu_N$. These values are quite small, which shows a good descriptive power of these observables in the present shell-model calculations.
   
   In Table \ref{BE2_BM1} we also present the shell-model predicted $B(E2)$ and $B(M1)$ transitions for those proton-hole-like states where at least a limit for the experimental data is available in any transition mode [$B(E2)$ or $B(M1)$].

\subsection{Configuration mixing in the single-hole-like states and spectroscopic factors}
\label{sec:sfactors}

\begin{figure}[b]
 	\includegraphics[width=85mm]{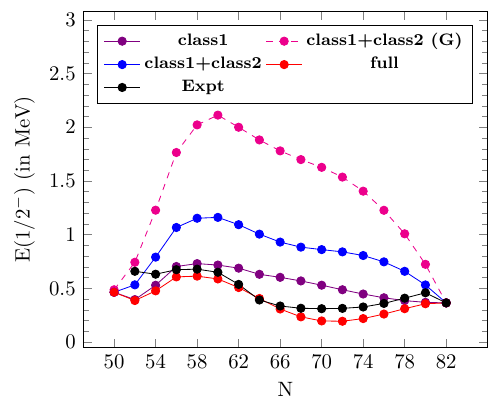}
 	\caption{Evolution of the experimental $1/2^-_1$ levels in odd-$A$ indium isotopes (Expt) compared to those of theory with different levels of configuration mixing included. The labels class 1, class 1 $+$ class 2, and full stand for the degrees of configuration mixing with the present effective interaction employed, and class 1 $+$ class 2 (G) indicates the class 1 $+$ class 2 model space with the $G$ matrix based proton-neutron interaction that was used in Ref. \cite{Patel}. See the text for details.}
 	\label{Energy_diff}
 \end{figure} 
 
Here, we carry out more detailed analyses of the configuration mixing caused by a proton hole for the $9/2^+_{\rm g.s.}$ and the $1/2^-_1$ levels. The dominant configurations of these states are naturally $[\pi(g_{9/2})^{-1}\times 0^+_1]^{(9/2^+)}$ and $[\pi(p_{1/2})^{-1}\times 0^+_1]^{(1/2^-)}$, respectively, for which $0^+_1$ stands for the ground state of the tin isotope with the same neutron number. {\color{black}Now,} we dub these configurations the class 1 configurations, because both the proton and the neutron wave functions have the lowest energies in their respective model spaces. In this context, the configurations with either the proton or the neutron ones excited from the lowest state can be regarded as the class 2 configurations. Typical excitation energies of protons and those of neutrons are both $\approx 1$~MeV or more. Since the proton excited states, i.e., $\pi(p_{3/2})^{-1}$ and $\pi(f_{5/2})^{-1}$, cannot produce $9/2^+$ or $1/2^-$ states by coupling to the $0^+_1$ state, the class 2 configurations consist only of the neutron excited states. The resulting class 2 configurations can be given by $[\pi(g_{9/2})^{-1}\times J^+_k]^{(9/2^+)}$ and $[\pi(p_{1/2})^{-1}\times J^+_k]^{(1/2^-)}$  ($J_k^+\ne 0^+_1$) for the $9/2^+_{\rm g.s.}$ and the $1/2^-_1$ states, respectively. It is then natural to define the class 3 configurations as those with both the proton and the neutron configurations excited from the lowest states, which include all the configurations in the present model space except the class 1 and class 2 configurations.

\begin{figure}
 	\includegraphics[width=85mm]{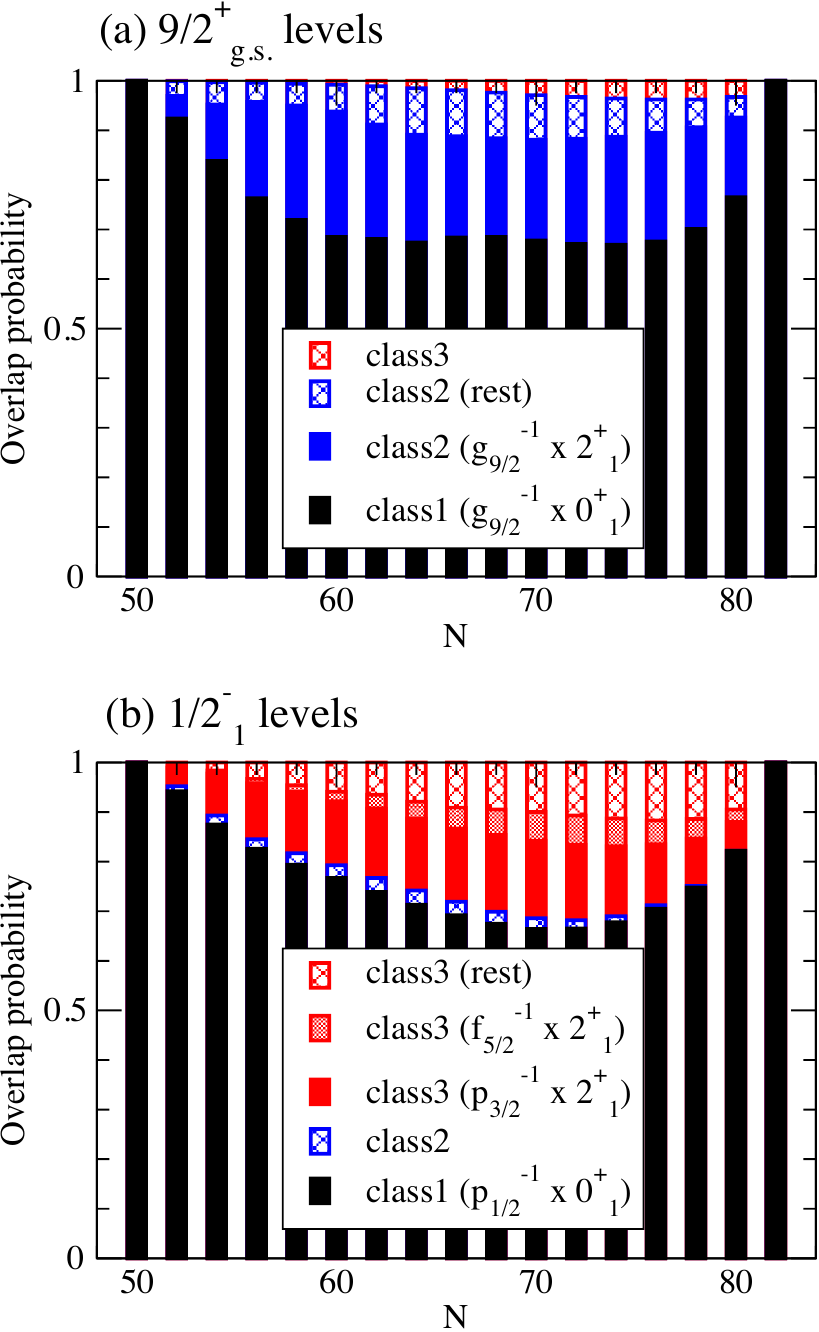}
 	\caption{Overlap probabilities of (a) the $9/2^+_{\rm g.s.}$ and (b) the $1/2^-_1$ states with a few dominant configurations. See the text for the notation of the configurations.}
 	\label{configuration}
 \end{figure} 

In the actual shell-model calculations, the energies of the class 1 configurations can be calculated by using two-body Hamiltonian matrix elements and the occupation numbers in the ground states of tin isotopes with the same neutron numbers, %as shown later. 
as we will present in formula (\ref{e_pn}) in Sec. \ref{sec:espe}. To evaluate the lowest energies in the class 1 $+$ class 2 configurations in a practical way, we perform shell-model calculations with a proton hole restricted to either the $p_{1/2}$ or the $g_{9/2}$ orbital. Note that some of the class 3 configurations, such as [$\pi(p_{1/2})^{-1}\times 5^-_1]^{9/2^+}$ for $9/2^+$, are admixed in this calculation but that their contributions are much smaller compared to the majority of the class 2 configurations. The class 3 configurations can be included by carrying out shell-model calculations in the full model space employed in this study. 

The effects of the higher order configurations on the $1/2^-_1$ levels in indium isotopes are shown in Fig. \ref{Energy_diff}. While the excitation energies in class 1 are rather constant and are not far from the experimental data along the isotope chain, those in class 1 $+$ class 2 increase for the neutron number away from the closed shells. As a result, a considerable deviation from the experimental levels arises. As denoted by class 1 $+$ class 2 (G) in Fig. \ref{Energy_diff}, this trend is similar but more enhanced with a $G$-matrix based proton-neutron interaction \cite{Patel} used in the same model space. This indicates that the deviation in class 1 $+$ class 2 is, aside from how much the energy increases, predominantly due to the degrees of restriction of the model space. By including the class 3 configurations in the full model space, the $1/2^-_1$ levels are lowered from those of class 1 $+$ class 2, and become very close to the experimental data on the whole. 

Here, we pursue why the $1/2^-_1$ levels behave as shown in Fig. \ref{Energy_diff} by enlarging the model space. First, we compare the $1/2^-_1$ energy levels in class 1 to those in class 1 $+$ class 2. As mentioned already, the class 2 configurations are $[\pi(g_{9/2})^{-1}\times J^+_k]^{(9/2^+)}$ and $[\pi(p_{1/2})^{-1}\times J^+_k]^{(1/2^-)}$ for the $9/2^+_{\rm g.s.}$ and the $1/2^-_1$ states, respectively. Here, the possible quantum number $J$ is limited to making a $9/2^+$ or $1/2^-$ state after the proton and the neutron angular momenta are coupled. More specifically, neutron states with $0 \le J \le 9$ are possible for $9/2^+$, whereas only those of $0 \le J \le 1$ are allowed for $1/2^-$. {\color{black}In particular, the coupling to the neutron state $2^+_1$, which is expected to be the largest fraction in class 2 because of the lowest excited state in tin isotopes, is missing in the latter (for $1/2^-$)}. As a result, the coupling to the class 2 lowers the energies of the $9/2^+_{\rm g.s.}$ state more than those of the $1/2^-_1$ state, thus increasing the $1/2^-_1$ excitation energies.

Next, we compare the $1/2^-_1$ energy levels in class 1 $+$ class 2 to those of the full configurations in which states with a proton hole in the $p_{3/2}$ or the $f_{5/2}$ orbital are also activated. The key to shifting the excitation energies of the $1/2^-_1$ states is parity coupling. For constructing the $9/2^+_{\rm g.s.}$ state, the coupling of the $p_{3/2}$ or the $f_{5/2}$ proton hole to a negative-parity state is allowed, but the coupling to a positive-parity state is prohibited. The situation for the $1/2^-_1$ state is the opposite. Since low-lying states in tin isotopes are dominated by positive-parity states, the coupling to class 3 configurations is larger for the $1/2^-_1$ states than for the $9/2^+_{\rm g.s.}$ state, decreasing the excitation energies of the $1/2^-_1$ states. 
 
As thus discussed, not only the class 2 configurations but also the class 3 configurations play a significant role in the $1/2^-_1$ excitation energies because the former and the latter configurations selectively favor the $9/2^+_{\rm g.s.}$ and the $1/2^-_1$ states, respectively. This situation is clearly seen in Fig. \ref{configuration}, in which dominant configurations in the $9/2^+_{\rm g.s.}$ and the $1/2^-_1$ states are contrasted. The $1/2^-_1$ excitation energies lowered by the class 3 configurations almost cancel those raised by the class 2 configurations. As a result of this process, the $1/2^-_1$ levels are located at the right positions. This fact is worth attention because shell-model calculations are often carried out in the proton valence shell that consists of the $p_{1/2}$ and $g_{9/2}$ orbitals and the neutron valence shell that includes the $d_{5/2}$, $g_{7/2}$, $s_{1/2}$, $d_{3/2}$, and $h_{11/2}$ orbitals \cite{Patel, Kast, Kownacki, Boelaert, Boelaert1, Patel1}, i.e., in class 1 $+$ class 2 in our notation.

\begin{table}
	 	\centering
	 	\caption{Comparison of the shell-model calculated spectroscopic factors ($C^2S$) for low-lying states of odd-$A$ indium isotopes with the experimental data \cite{Weiffenbach, Blachot, Blachot1, Blachot2} from the $^{A+1}$Sn($d,^3$He)$^A$In reactions.}
	 	\begin{ruledtabular}
	 		\begin{tabular}{ccccccc}
	 			%\hline \hline
	 			
	 			Isotope & $J^{\pi}$	& \makecell{Theory \\ (keV)} & \makecell{Expt. \\ (keV)} &	$L$ & \multicolumn{2}{c}{$C^2S$}	 \\
	 			
	 			\cline{6-7}
	 			
	 			&  &  &  &  & SM & Expt. \\
	 			
	 			\hline \\

     $^{99}$In & $9/2^+_{\rm g.s.}$ & 0 & - & 4 & 10.000 & - \\
	 			        
	 			        & $1/2^-_1$   & 464 & - & 1  & 2.000 & - \\
	 			        
	 			        & $3/2^-_1$   & 1339 & - & 1  & 4.000 & - \\
	 			        
	 			        & $5/2^-_1$   & 3261 & - & 3  & 6.000 & - \\

                        $^{101}$In & $9/2^+_{\rm g.s.}$  & 0  & 0  & 4  & 9.243 & - \\

                        & $1/2^-_1$   & 387 & 659 & 1  & 1.884 & - \\

                        & $3/2^-_1$   & 1164 & - & 1  & 3.388 & - \\

                        & $5/2^-_1$   & 1934 & - & 3  & 0.094 & - \\

                        $^{103}$In & $9/2^+_{\rm g.s.}$  & 0  & 0  & 4  & 8.376 & - \\

                        & $1/2^-_1$   & 477 & 632 & 1  & 1.751 & - \\

                        & $3/2^-_1$   & 1085 & - & 1  & 2.849 & - \\

                        & $5/2^-_1$   & 1817 & - & 3  & 0.114 & - \\

                        $^{105}$In & $9/2^+_{\rm g.s.}$  & 0  & 0  & 4  & 7.628 & - \\

                        & $1/2^-_1$   & 607 & 674 & 1  & 1.652 & - \\

                        & $3/2^-_1$   & 1094 & - & 1  & 2.652 & - \\

                        & $5/2^-_1$   & 1828 & - & 3  & 0.163 & - \\

                        $^{107}$In & $9/2^+_{\rm g.s.}$  & 0  & 0  & 4  & 7.193 & - \\

                        & $1/2^-_1$   & 614 & 679 & 1  & 1.588 & - \\

                        & $3/2^-_1$   & 1061 & 1107 & 1  & 2.554 & - \\

                        & $5/2^-_1$   & 1770 & 1519 & 3  & 0.255 & - \\

                        $^{109}$In & $9/2^+_{\rm g.s.}$  & 0  & 0  & 4  & 6.848 & - \\

                        & $1/2^-_1$   & 588 & 650 & 1  & 1.534 & - \\

                        & $3/2^-_1$   & 1021 & 981 & 1  & 2.436 & - \\

                        & $5/2^-_1$   & 1705 & 1441 & 3  & 0.456 & - \\

	 			$^{111}$In & $9/2^+_{\rm g.s.}$  & 0  & 0  & 4  & 6.815 & 5.5 \\
	 			
	 			& $1/2^-_1$  & 508 & 537 & 1  & 1.478 & 1.5 \\
	 			
	 			& $3/2^-_1$  & 934 & 803 & 1  & 2.239 & 2 \\

                        & $5/2^-_1$  & 1585 & 1280 & 3  & 0.739 & - \\

	 			$^{113}$In & $9/2^+_{\rm g.s.}$  & 0  & 0  & 4  & 6.746 & 6.0 \\
	 			
	 			& $1/2^-_1$  & 406 & 392 & 1  & 1.426 & 1.3 \\
	 			
	 			& $3/2^-_1$  & 847 & 647 & 1  & 1.997 & 1.7 \\

                        & $5/2^-_1$  & 1400 & 1106 & 3  & 0.738 & - \\
	 			
	 			$^{115}$In & $9/2^+_{\rm g.s.}$  & 0  & 0  & 4  & 6.844 & 6.7 \\
	 			
	 			& $1/2^-_1$  & 308 & 336 & 1  & 1.384 & 1.5 \\
	 			
	 			& $3/2^-_1$  & 772 & 597 & 1  & 1.767 & 1.9 \\
	 			
	 			& $5/2^-_1$  & 1230 & 1041 & 3  & 0.693 & 0.7 \\
	 			
	 			$^{117}$In & $9/2^+_{\rm g.s.}$  & 0  & 0  & 4  & 6.852 & 6.7 \\
	 			
	 			& $1/2^-_1$  & 235 & 315 & 1  & 1.349 & 1.5 \\
	 			
	 			& $3/2^-_1$  & 710 & 589 & 1  & 1.608 & 2.3 \\

                        & $5/2^-_1$  & 1103 & 1028 & 3  & 0.655 & - \\
	 			
	 			$^{119}$In & $9/2^+_{\rm g.s.}$  & 0  & 0  & 4  & 6.785 & 6.5 \\
	 			
	 			& $1/2^-_1$  & 197 & 311 & 1  & 1.329 & 1.6 \\
	 			
	 			& $3/2^-_1$  & 678 & 604 & 1  & 1.526 & 1.8 \\

                        & $5/2^-_1$  & 1033 & 1044 & 3  & 0.622 & - \\
	 			
	 			$^{121}$In & $9/2^+_{\rm g.s.}$  & 0  & 0  & 4  & 6.712 & 7.2 \\
	 			
	 			& $1/2^-_1$  & 194 & 314 & 1  & 1.329 & 1.4 \\
	 			
	 			& $3/2^-_1$  & 686 & 638 & 1  & 1.513 & 2.1 \\

                        & $5/2^-_1$  & 1029 & 1079 & 3  & 0.597 & - \\
	 			
	 			$^{123}$In & $9/2^+_{\rm g.s.}$  & 0  & 0  & 4  & 6.685 & 7.2 \\
	 			
	 			& $1/2^-_1$  & 219 & 327 & 1  & 1.356 & 1.4 \\
	 			
	 			& $3/2^-_1$  & 736 & 699 & 1  & 1.572 & 1.7 \\

                        & $5/2^-_1$  & 1087 & 1138 & 3  & 0.581 & - \\

     \end{tabular}
	 	\end{ruledtabular}
	 	\label{spectroscopic_factor}
	 \end{table}
 \addtocounter{table}{-1}
\begin{table}[h]
%	 	\centering
%	 	TABLE II. (Continued)
\caption{(Continued)}
	 	\begin{ruledtabular}
	 		\begin{tabular}{ccccccc}
	 			%\hline \hline
	 			
	 			Isotope & $J^{\pi}$	& \makecell{Theory \\ (keV)} & \makecell{Expt. \\ (keV)} &	$L$ & \multicolumn{2}{c}{$C^2S$}	 \\
	 			
	 			\cline{6-7}
	 			
	 			&  &  &  &  & SM & Expt. \\
	 			
	 			\hline \\

                        $^{125}$In & $9/2^+_{\rm g.s.}$  & 0  & 0  & 4  & 6.763 & - \\
                        
                        & $1/2^-_1$  & 261 & 360 & 1  & 1.410 & - \\
                        
                        & $3/2^-_1$  & 821 & 796 & 1  & 1.699 & - \\
                        
                        & $5/2^-_1$  & 1193 & 1220 & 3  & 0.564 & - \\

                        $^{127}$In & $9/2^+_{\rm g.s.}$  & 0  & 0  & 4  & 7.010 & - \\

                        & $1/2^-_1$  & 311 & 409 & 1  & 1.495 & - \\

                        & $3/2^-_1$  & 936 & 933 & 1  & 1.883 & - \\

                        & $5/2^-_1$  & 1315 & - & 3  & 0.520 & - \\

                        $^{129}$In & $9/2^+_{\rm g.s.}$  & 0  & 0  & 4  & 7.655 & - \\

                        & $1/2^-_1$  & 357 & 459 & 1  & 1.643 & - \\

                        & $3/2^-_1$  & 1089 & 1091 & 1  & 2.191 & - \\

                        & $5/2^-_1$  & 1407 & - & 3  & 0.425 & - \\

                        $^{131}$In & $9/2^+_{\rm g.s.}$  & 0  & 0  & 4  & 10.000 & - \\

                        & $1/2^-_1$  & 365 & 365 & 1  & 2.000 & - \\

                        & $3/2^-_1$  & 1352 & 1353 & 1  & 4.000 & - \\

                        & $5/2^-_1$  & 3275 & 3275 & 3  & 6.000 & - \\

	 		\end{tabular}
	 	\end{ruledtabular}
	 	%\label{spectroscopic_factor2}
	 \end{table}

    \begin{figure*}
   	\includegraphics[width=155mm]{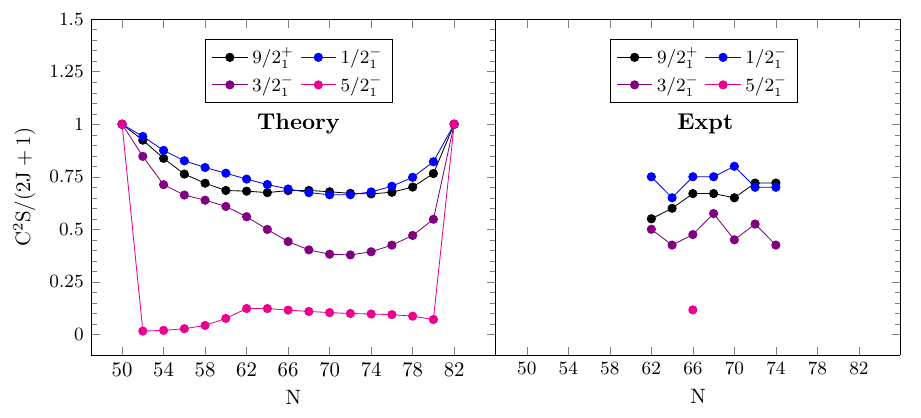}
   	\caption{Comparison of the shell-model predicted and experimental spectroscopic factors $(C^2S)$ divided by $2J+1$ for $9/2^+_{\rm g.s.}$, $1/2^-_1$, $3/2^-_1$, and $5/2^-_1$ states.}
   	\label{spectro}
   \end{figure*}

The overlap probabilities between an eigenstate and a specific configuration, such as those presented in Fig. \ref{configuration}, can be probed with the spectroscopic factor that is deduced from nucleon transfer or knockout reactions. Here, an eigenstate with the mass number $A$ is denoted as $|\alpha\rangle=| \Psi^A \omega JM \rangle$, where $J$ and $M$ are the total angular momentum and its $z$ component, respectively, and $\omega$ is introduced to label the state with the same $(J,M)$. The hole state with the same quantum numbers is expressed as $|\beta\rangle = \mathcal{N}\left[ \tilde{c}_{j} \times | \Psi^{A+1} \omega' J' M' \rangle \right]^{(J)}_M$, where $\mathcal{N}$ is the normalization factor, and $\tilde{c}_{jm} = (-1)^{j+m}c_{j\,-m}$ is the hole creation operator. The overlap between $|\alpha\rangle$ and $|\beta\rangle$ is represented as
\begin{equation}
\begin{array}{rcl}
 \langle \beta | \alpha \rangle  &=& \mathcal{N} \sum_{mM'}(jmJ'M'|JM) (-1)^{j+m} \\
& & \times \langle  \Psi^{A+1} \omega' J'M' | c^{\dagger}_{j\,-m} |  \Psi^A \omega JM \rangle \\
&=& \mathcal{N} \sum_{mM'} \sqrt{2J+1} \left(
\begin{array}{ccc}
j & J' & J \\
m & M' & -M \\
\end{array}
\right)^2 \\
& & \times  \langle \Psi^{A+1} \omega' J' \lVert c^{\dagger}_{j} \rVert \Psi^A \omega J \rangle \\
&=& \displaystyle \mathcal{N}\frac{\langle \Psi^{A+1} \omega' J' \lVert c^{\dagger}_{j} \rVert \Psi^A \omega J \rangle}{\sqrt{2J+1}}.
\end{array}
\end{equation}
The corresponding overlap probability is expressed as
\begin{equation}
\lvert \langle \beta | \alpha \rangle \rvert^2 = 
\mathcal{N}^2 \displaystyle\frac{2J'+1}{2J+1}C^2S_j(\omega J,\omega' J'), 
\label{eq:overlap_prob}
\end{equation}
using the spectroscopic factor defined as
\begin{equation}
    C^2S_j(\omega J,\omega' J') = \displaystyle \frac{\lvert \langle \Psi^{A+1} \omega' J' \lVert c^{\dagger}_{j} \rVert \Psi^A \omega J \rangle \rvert^2}{2J'+1} , 
\end{equation}
where $\omega J$ and $\omega' J'$ are used to designate the initial and the final states we take. If not necessary, these labels are omitted, and the spectroscopic factors are then simply denoted as $C^2S_j$ or $C^2S$. In the present study, we consider the situation in which a proton hole is created by $\tilde{c}_{jm}$ ($j=g_{9/2}, p_{1/2}, p_{3/2}, f_{5/2}$) on top of the fully occupied states, $\pi(f_{5/2})^{6}(p_{3/2})^{4}(p_{1/2})^{2}(g_{9/2})^{10}$. In such a case, the normalization factor $\mathcal{N}$ is always unity, and the overlap probability given by Eq. (\ref{eq:overlap_prob}) can directly be deduced from the spectroscopic factor. The cumulative sum of the overlap probabilities of the class 1 and class 2 configurations is the sum of $\lvert \langle \beta | \alpha \rangle \rvert^2$ over all the states in the tin isotopes with the hole state $j$ fixed (to $\pi g_{9/2}$ for the $9/2^+_1$ level or to $\pi p_{1/2}$ for the $1/2^-_1$ level). Hence for $\mathcal{N}=1$ it reads
\begin{equation}
\begin{array}{rcl}
\sum _{\omega' J'} \lvert \langle \beta | \alpha \rangle \rvert^2 & = &  \sum _{\omega' J'} \displaystyle\frac{2J'+1}{2J+1}C^2S_j(\omega J,\omega' J') \\
 &=& (2j+1) - n_j(\alpha) ,
\end{array}
\end{equation}
where $n_j(\alpha)$ stands for the nucleon occupation number in the orbital $j$ for the state $|\alpha\rangle$. The overlap probabilities plotted in Fig. \ref{configuration} are thus obtained with the $C^2S$ and the $n_j(\alpha)$ values in the kshell outputs.

The experimental $C^2S$ values for $9/2^+_{\rm g.s.}$, $1/2^-_1$, $3/2^-_1$, and $5/2^-_1$ states are available from the one-proton pickup reaction $^{A+1}$Sn($d,^3$He)$^A$In for stable tin isotopes. The results of the theoretical and experimental $C^2S$ values are summarized in Table \ref{spectroscopic_factor}. The experimental values for the $9/2^+_{\rm g.s.}$, $1/2^-_1$, and $3/2^-_1$ states in $^{115-123}$In, $^{111}$In, and $^{113}$In are taken from Refs. \cite{Weiffenbach}, \cite{Blachot}, and \cite{Blachot1}, respectively. The experimental value for the $5/2^-_1$ state in $^{115}$In is taken from Ref. \cite{Blachot2}.

  The calculated $C^2S$ values for the $9/2^+_{\rm g.s.}$, $1/2^-_1$, $3/2^-_1$, and $5/2^-_1$ states show remarkable agreement with the available experimental data. To emphasize to what extent the single-hole nature is kept in these states, we plot in Fig. \ref{spectro} these $C^2S$ values divided by $2J+1$, which are identical with the overlap probabilities of the class 1 configuration, as shown in Eq. (\ref{eq:overlap_prob}) (note that $J'=0$). It is known that experimental spectroscopic factors, especially the overall factor, cannot be free from some systematic uncertainties because deducing the spectroscopic factors relies on reaction theories. As seen in Fig. \ref{spectro}, the experimental $C^2S$ values stagger with neutron number compared to smooth evolutions predicted by shell-model calculations. It is most likely that this staggering is due to the experimental uncertainties, although the experimental error bars are not provided. By considering the experimental uncertainties, the shell model well captures what the experimental data tell us: (i) The $C^2S/(2J+1)$ values for the $9/2^+_{\rm g.s.}$ and the $1/2^-_1$ states are about 0.7 in mid-shell, pointing to significant configuration mixing even for the lowest two states. (ii) The $C^2S/(2J+1)$ values for $3/2^-_1$ are reduced to $\approx 0.5$, indicating further depletion of the single-hole strengths. (iii) The $C^2S/(2J+1)$ value for $5/2^-_1$ in $^{115}$In is only $\approx 0.1$, suggesting the dominance of a collective state, $\pi(p_{1/2})^{-1}\times 2^+_1$. \\

\subsection{Effective single-hole energies}
\label{sec:espe}

While significant configuration mixing occurs in the $9/2^+_{\rm g.s.}$, $1/2^-_1$, $3/2^-_1$, and $5/2^-_1$ states as seen already, proton single-hole characters still remain, particularly in the $9/2^+_{\rm g.s.}$ and the $1/2^-_1$ states. Hence, it is worth examining the systematics of these energies in terms of effective single-particle energies (ESPEs).

For a fully occupied orbital $j$, the ESPE is defined as the energy that is needed to add a nucleon hole in the orbital $j$ by using the monopole interaction (see Sec. III, pp $7-11$ of Ref. \cite{Otsuka1}). In the present case, we consider the $Z=50$ core, for which the effective proton single-hole energies  $\epsilon({j_p^{-1}})$ for a state with $n_{j_n}$ neutrons occupying the orbitals $j_n$ are given by 
\begin{equation}
\begin{array}{rcl}
\epsilon({j_p^{-1}}) & = & \epsilon^0({j_p^{-1}}) + \sum_{j_n} V^m_{pn}(j_p^{-1}, j_n) n_{j_n}  \\
     & = & \epsilon^0({j_p^{-1}}) - \sum_{j_n} V^m_{pn}(j_p, j_n) n_{j_n} , \\
\end{array}
\label{espe}
\end{equation}
where $V^m_{pn}(j_p, j_n)$ is the monopole matrix element between $j_p$ and $j_n$ given by 
\begin{equation}
\begin{array}{rcl}
     V^m_{pn}(j_p, j_n) &=& \displaystyle \frac{\sum_{J=|j_p-j_n|}^{j_p+j_n}(2J+1)V_J(j_p j_n j_p j_n)}{\sum_{J=|j_p-j_n|}^{j_p+j_n} (2J+1)} \\
     &=& \displaystyle \frac{\sum_{J=|j_p-j_n|}^{j_p+j_n}(2J+1)V_J(j_p j_n j_p j_n)}{(2j_p+1)(2j_n+1)} , 
\end{array}
\label{monopole_mat}
\end{equation}
and $\epsilon^0({j_p^{-1}})$ is the bare single-hole energy of {\color{black}orbital} $j_p$. The diagonal matrix element of the particle-hole interaction is represented by using the Pandya transformation \cite{Pandya} as 
\begin{equation}
\begin{array}{rcl}
& & V_J(j_p^{-1} j_n j_p^{-1} j_n) \\
&=& -\sum_{J'}(2J'+1) \left\{
\begin{array}{ccc}
   j_p  & j_n & J' \\
   j_p  & j_n & J
\end{array}
\right\} V_{J'}(j_p j_n j_p j_n) , 
\end{array}
\end{equation}
from which $V^m_{pn}(j_p^{-1}, j_n) = - V^m_{pn}(j_p, j_n)$ is derived, and is used in Eq. (\ref{espe}). 

To obtain effective proton single-particle (or single-hole) energies, one often assumes that each neutron orbital $j_n$ 
is either completely vacant or completely occupied, i.e., $ n_n=0$ or $ n_n =2j_n+1$. Here, we dub this conventional effective single-particle energy ESPE1. However, Eq. (\ref{espe}) can also be applied to fractional neutron occupancies. The natural choice for defining such generalized effective proton single-hole energies, which we call ESPE2, is to use the neutron occupancies of the $0^+_{\rm g.s.}$ state of the tin isotope with the same neutron number. 

\begin{figure}
	\includegraphics[width=85mm]{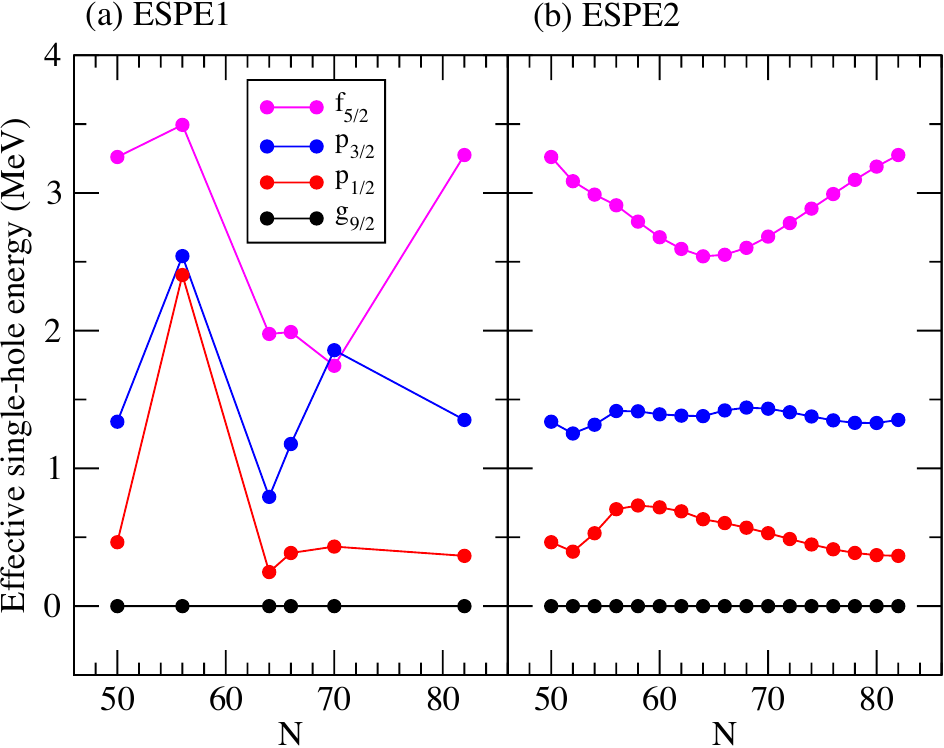}
	\caption{ {\color{black}Effective proton} single-hole energies of the $g_{9/2}$, $p_{1/2}$, $p_{3/2}$, and $f_{5/2}$ orbitals with the definitions of (a) ESPE1 and (b) ESPE2, measured from those of $g_{9/2}$. For ESPE1, it is assumed that neutrons fill the orbitals in the order of $d_{5/2}$, $g_{7/2}$, $s_{1/2}$, $d_{3/2}$, and $h_{11/2}$.  
 }
	\label{eff_spe}
	\end{figure} 
 
\begin{figure}
	\includegraphics[width=85mm]{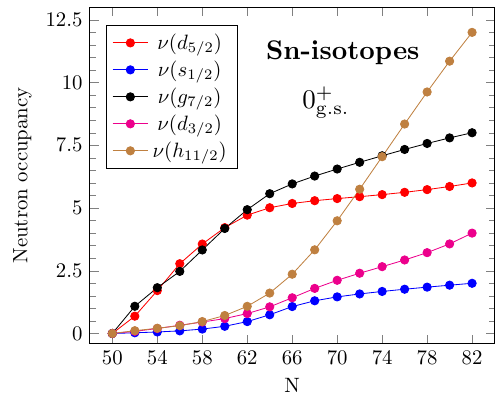}
	\caption{
 Neutron occupancy for the $0_{\rm g.s.}^+$ state in even-$A$ tin isotopes.}
	\label{Occupancy}
	\end{figure}

The effective proton single-hole energies measured from $g_{9/2}$ orbital are illustrated in Fig. \ref{eff_spe}. Here, we focus on the evolution of excitation energies of the proton hole in the $p_{1/2}$ orbital. When the orbital $j_n$ is filled by $n_{j_n}$ neutrons, the excitation energy changes by 
\begin{equation}
    \Delta = \left( V^m_{pn} (g_{9/2},j_n) - V^m_{pn} (p_{1/2},j_n) \right) n_{j_n} .
\end{equation}
As discussed in Sec. IV, pp $14-30$ of Ref. \cite{Otsuka1}, the central and the tensor forces give major contributions to the monopole matrix elements. The central force produces a large negative monopole matrix element between two orbitals $(n_1l_1j_1)$ and $(n_2l_2j_2)$ for $n_1 = n_2$, and the latter gives attraction when $(j_1, j_2)=(l_1+1/2, l_2-1/2)$ or $(j_1, j_2)=(l_1-1/2, l_2+1/2)$ is satisfied. These general properties are applied to the neutron and the proton orbitals of interest, resulting in the labels summarized in Table \ref{monopole}. From these labels, one easily finds that the $p_{1/2}$ proton hole excitation energy sharply {\color{black}increases and decreases} when neutrons fill the $d_{5/2}$ and $g_{7/2}$ orbitals, respectively. The other neutron orbitals {\color{black}($s_{1/2}$, $d_{3/2}$, and $h_{11/2}$)} produce similar effects on the $g_{9/2}$ and the $p_{1/2}$ proton orbitals. The $p_{1/2}$ proton hole excitation energy is kept almost constant when these neutron orbitals are filled. The behavior of the $p_{1/2}$ proton hole excitation energy presented in Fig. \ref{eff_spe}(a) is thus understood. 

\begin{table}
	 	\centering
\caption{ Labeling several pairs of the orbitals $(j_p,j_n)$ of interest according to the strengths of the monopole matrix elements $V_{pn}^m(j_p, j_n)$ by the central and the tensor forces. The labels ``f", ``u", and ``0" stand for the favored (most attractive), unfavored (less attractive or repulsive), and vanishing matrix elements, respectively. The first and the second labels are for the central and the tensor forces, respectively. } 
	 	\begin{ruledtabular}
	 		\begin{tabular}{cccccc}
    $j_p \backslash j_n$  & $1d_{5/2}$ & $0g_{7/2}$ & $2s_{1/2}$ & $1d_{3/2}$ & $0h_{11/2}$   \\ \hline
     $0g_{9/2}$ & (u, u) & (f, f) & (u, 0) & (u, f) & (f, u) \\
     $1p_{1/2}$ & (f, f) & (u, u) & (u, 0) & (f, u) & (u, f) 
	 		\end{tabular}
	 	\end{ruledtabular}
	 	\label{monopole}
\end{table}

The effective proton single-hole energies defined as ESPE2 are depicted in Fig. \ref{eff_spe}(b). Here, we can notice that the sharp peak at $N=56$ seen in Fig. \ref{eff_spe}(a) disappears, and the $p_{1/2}$ proton hole excitation energy change is much milder. This is due to the fact that we are considering here the actual neutron occupancies shown in Fig. \ref{Occupancy}: Neutrons predominantly fill the $d_{5/2}$ and $g_{7/2}$ orbitals almost equally for $N \lesssim 64$ because of their similar single-particle energies and a large pairing matrix element of $V_{J=0}(d_{5/2} d_{5/2} g_{7/2} g_{7/2})$, and then fill the remaining orbitals for $N \gtrsim 64$. As a result, the opposite effects of the neutron $d_{5/2}$ and $g_{7/2}$ orbitals on the proton $g_{9/2}$ and $p_{1/2}$ orbitals are almost canceled out. 

In the following, we prove that the excitation energies due to ESPE2 are identical with those of the class 1 configuration of the corresponding orbitals, as shown in Fig. \ref{Energy_diff}. To calculate the energy of the class 1 configuration, the proton hole state is expressed as 
\begin{equation}
    \Phi_{j_p m_p} = d^{\dagger}_{j_pm_p}\Psi(0^+_1), 
\end{equation}
where $d^{\dagger}_{j_pm_p} = \tilde{c}_{j_p m_p} = (-1)^{j_p+m_p}c_{j_p\,-m_p}$ is the proton-hole creation operator and $\Psi(0^+_1)$ is the ground state of the tin core. The expectation value of the Hamiltonian for $\Phi_{j_p m_p}$ is written as 
\begin{equation}
\begin{array}{rcl}
  E(j_p m_p) & =& \langle \Phi_{j_p m_p} | H | \Phi_{j_p m_p} \rangle \\
  & =& \epsilon^0({j_p^{-1}}) + \langle \Phi_{j_p m_p} | V_{pn} | \Phi_{j_p m_p} \rangle \\ 
  & & + \langle \Psi(0^+_1) | H_{nn}| \Psi(0^+_1) \rangle ,  \\
\end{array}
\label{energy_hole}
\end{equation}
where $V_{pn}$ and $H_{nn}$ are the proton-neutron interaction and the neutron sectors of the Hamiltonian, respectively. From Eq. (\ref{twobody}), $V_{pn}$ is expressed as
\begin{equation}
\begin{array}{rcl}
     V_{pn} &=& \sum_{j'_p j''_p j'_n j''_n m'_p m''_p m'_n m''_n JM} V_J\left( (j'_p)^{-1} j'_n (j''_p)^{-1} j''_n \right) \\
     & & \times (j'_p m'_p j'_n m'_n | JM) (j''_p m''_p j''_n m''_n | JM) \\
     & & \times d^{\dagger}_{j'_p m'_p} c^{\dagger}_{j'_n m'_n} c_{j''_n m''_n} d_{j''_p m''_p},
\end{array}
\end{equation}
with the particle-hole matrix elements $V_J\left((j'_p)^{-1} j'_n (j''_p)^{-1} j''_n \right)$. Its expectation value for $\Phi_{j_p m_p}$ is calculated to be 
\begin{equation}
\begin{array}{rcl}
  & &  \langle \Phi_{j_p m_p} | V_{pn} | \Phi_{j_p m_p} \rangle \\
    & = & \sum_{m_p} \displaystyle \frac{\langle \Phi_{j_p m_p} | V_{pn} | \Phi_{j_p m_p} \rangle}{2j_p+1} \\
    & = & \sum_{j'_n m'_n JM m_p} \displaystyle \frac{V_J( j_p^{-1} j'_n j_p^{-1} j'_n)}{2j_p+1} \\
    &   & \times (j_p m_p j'_n m'_n | JM)^2 \langle \Psi (0^+_1) | c^{\dagger}_{j'_n m'_n} c_{j'_n m'_n} | \Psi (0^+_1) \rangle \\
    & = &  \sum_{J j'_n } \displaystyle \frac{(2J+1)V_J(j_p^{-1} j'_n j_p^{-1} j'_n )}{(2j_p+1)(2j'_n+1)} n_{j'_n} \\
    & = & \sum_{j'_n } V_{pn}^{m}(j_p^{-1},j'_n) n_{j'_n}  \\
    & = & -\sum_{j'_n } V_{pn}^{m}(j_p,j'_n) n_{j'_n},
\end{array}
\label{pn_energy}
\end{equation}
with $n_{j'_n} = \langle \Psi (0^+_1) | \sum_{m'_n} c^{\dagger}_{j'_n m'_n} c_{j'_n m'_n} | \Psi (0^+_1) \rangle$. {\color{black}The first equality of Eq. (\ref{pn_energy}) expresses the fact that the energy does not depend on $m_p$.} The second equality is obtained by assuming that there is no pair of the neutron orbitals with the same angular momentum and parity in the valence shell, such as $1d_{5/2}$ and $2d_{5/2}$, resulting in $\langle \Psi (0^+_1) | c^{\dagger}_{j'_n m'_n} c_{j''_n m''_n} | \Psi (0^+_1) \rangle = \delta_{j'_n j''_n}\delta_{m'_n m''_n} \langle \Psi (0^+_1) | c^{\dagger}_{j'_n m'_n} c_{j'_n m'_n} | \Psi (0^+_1) \rangle$, 
and by using $\langle - | d_{j_p m_p} d^{\dagger}_{j'_p m'_p} d_{j''_p m''_p} d^{\dagger}_{j_p m_p} | - \rangle = \delta_{j_pj'_p}\delta_{j_pj''_p}\delta_{m_pm'_p}\delta_{m_pm''_p}$ for the vacuum {\color{black}state} $|-\rangle$. The third equality follows the orthogonality of the Clebsch-Gordan coefficient, $\sum_{M m_p} (j_p m_p j'_n m'_n | JM)^2= (2J+1)/(2j'_n+1)$. {\color{black} The fourth equality follows from Eq. (\ref{monopole_mat}).}

Hence, Eq. (\ref{energy_hole}) {\color{black}can be} written as
\begin{equation}
    E(j_p m_p) = \epsilon^0({j_p^{-1}}) -\sum_{j'_n } V_{pn}^{m}(j_p,j'_n)  n_{j'_n} + E_{nn},
    \label{e_pn}
\end{equation}
with $E_{nn}=\langle \Psi(0^+_1) | H_{nn}| \Psi(0^+_1) \rangle$. Since $E_{nn}$ does not depend on the proton hole state $(j_p,m_p)$, we do not have to consider it when only the relative energy, $E(j_p m_p) - E(j'_p m'_p)$, is of interest. As a result, Eq. (\ref{e_pn}) is the same as Eq. (\ref{espe}) when its neutron part is ignored.  

In the way we demonstrated above, the evolution of the $1/2^-_1$ excitation energies in the class 1 configuration, which is identical with the $p_{1/2}$ proton hole excitation energies in ESPE2, should be rather mild because of the almost equal filling of neutrons in $d_{5/2}$ and $g_{7/2}$ orbitals for the $0^+_{\rm g.s.}$ state in the lower mass tin isotopes. A very similar trend for the $p_{1/2}$ proton hole excitation energies in ESPE2 is also obtained if the proton-neutron interaction is replaced with the $G$-matrix based one that was used in Ref. \cite{Patel}. It is thus most likely that the difference between class 1 $+$ class 2 and class 1 $+$ class 2 (G) in Fig. \ref{Energy_diff} occurs predominantly because the coupling energies of the class 1 and the class 2 configurations is larger with the $G$-matrix interaction than with the $V_{\rm MU}$ interaction. The $1/2^-_1$ energy levels must be more or less peaked independently of the interaction in the class 1 $+$ class 2 configurations, as seen in Fig. \ref{Energy_diff}, and the experimental excitation energies can only be reproduced with the class 3 configurations included, i.e., with shell-model calculations including the $\pi p_{3/2}$ and the $\pi f_{5/2}$ orbitals as the valence shell.

	\section{Summary and Conclusions} \label{section4}

    We have investigated the modification of single-hole-like states by configuration mixing in the $^{99-131}$In isotopes with large-scale shell-model calculations in the proton and neutron model spaces of $Z=28-50$ and $N=50-82$, respectively. The proton-neutron effective interaction, which is essential in shell evolution, is taken from a variant of the $V_{\textrm{MU}}$ interaction that was previously employed for a consistent description of the proton-particle-like states in $^{101-133}$Sb. The calculated energy levels of the $1/2^-_1$, $3/2^-_1$, and $5/2^-_1$ states with respect to the $9/2^+_{\rm g.s.}$ state are in remarkable agreement with the experimental data. The evolution of the electromagnetic moments in the $9/2^+_{\rm g.s.}$ and the $1/2^-_1$ levels is also well reproduced, indicating significant configuration mixing in these states for the neutron number away from the closed shells.
    
    To gain more insight into the configuration mixing in these levels, the wave functions are decomposed into several dominant configurations that are classified according to the order of energy. The single-hole strengths of the $9/2^+_{\rm g.s.}$ and the $1/2^-_1$ states are both reduced to $\approx 0.7$ toward mid-shell, which is supported by the measured spectroscopic factors taken from the $^{A+1}$Sn($d$, $^3$He)$^A$In reactions. On the other hand, the remaining configurations show unique features: (i) For the class 2 configurations in terms of energy, the $\pi (g_{9/2})^{-1}\times 2^+_1$ configuration occupies a large fraction of the $9/2^+_{\rm g.s.}$ state, but the corresponding configuration in the $1/2^-_1$ states, i.e., $\pi (p_{1/2})^{-1}\times 2^+_1$, cannot make $J=1/2$ and is therefore excluded. (ii) The $\pi (p_{3/2})^{-1}\times 2^+_1$ and $\pi (f_{5/2})^{-1}\times 2^+_1$ configurations can be mixed with the negative-parity states, but are excluded from admixture in the $9/2^+_{\rm g.s.}$ state due to parity conservation. Although the effects of case (ii) are regarded as class 3 in terms of energy, its energy gain is comparable to that of (i). As a result, the effects of (i) and (ii) on the $1/2^-_1$ excitation energies are almost canceled out, which makes these excitation energies close to the effective proton single-hole energies. The $\pi p_{1/2}$ effective single-hole energies measured from those of $\pi g_{9/2}$ should change rather mildly along $N=50-82$, following the general properties of the monopole matrix elements of the central and tensor forces, and the nearly equal filling of the $\nu d_{5/2}$ and $\nu g_{7/2}$ orbitals for the lower neutron numbers. {\color{black}The evolution of the $1/2^-_1$ levels is thus understood in terms of the configuration mixing.} In particular, it is worth pointing out that these energies should deviate from the {\color{black}experimental data without the inclusion of} $\pi p_{3/2}$-$\pi f_{5/2}$ orbitals because of the missing correlation energy due to (ii). \\

	\section*{{Acknowledgments}}
	
	We acknowledge financial support from MHRD, the Government of India, and SERB (India), Grant No. CRG/2022/005167. We would like to thank the National Supercomputing Mission (NSM) for providing computing resources of ``PARAM Ganga" at the Indian Institute of Technology Roorkee, implemented by C-DAC and supported by MeitY and DST, Government of India. N.S. and Y.U. acknowledge the support of the ``Program for promoting researches on the supercomputer Fugaku," MEXT, Japan (Grant No. JPMXP1020230411), and the support of JSPS KANENHI Grant No. 20K03981. N.S. acknowledges the MCRP program of the Center for Computational Sciences, University of Tsukuba (NUCLSM).

\end{document}